\documentclass{IEEEtran}
\usepackage{cite}
\usepackage{amsmath,amssymb,amsfonts}
\usepackage{graphicx}
\usepackage{textcomp,nicefrac}
\usepackage{float}
\usepackage{subfig}
\usepackage[hyphens]{url}

\def\BibTeX{{\rm B\kern-.05em{\sc i\kern-.025em b}\kern-.08em
T\kern-.1667em\lower.7ex\hbox{E}\kern-.125emX}}
\newcommand\blfootnote[1]{%
  \begingroup
  \renewcommand\thefootnote{}\footnote{#1}%
  \addtocounter{footnote}{-1}%
  \endgroup
}

\begin{document}
\title{Cooling and Timing Tests of the ATLAS Fast Tracker VME boards}
\author{S. Sottocornola, A. Annovi, N.V. Biesuz, E. Brost, M. Calvetti, C. Gentsos, T. Holmes, L. Horyn,  
T. Iizawa, \mbox{A. Lanza}, J.D. Long, P. Mastrandrea, I. Maznas, A. Negri, D. Calabro, M. Piendibene, C. Roda, E. Romano, \mbox{T. Seiss}.
\thanks{
S. Sottocornola, A. Lanza, A. Negri, D. Calabro, E. Romano are with Sezione di Pavia INFN, via Bassi 6, Pavia, Italy 
(email: simone.sottocornola@cern.ch).
A. Annovi, N. V. Biesuz, P. Mastrandrea, M. Calvetti, M. Piendibene and C. Roda are with the University of Pisa and INFN Sezione di Pisa.
E. Brost is with Brookhaven National Lab.
C. Gentsos is with CERN.
T. Holmes, L. Horyn, T. Seiss are with University of Chicago.
J. D. Long is with University of Illinois Urbana-Champaign.
T.Iizawa is with University of Geneva.
I. Maznas is with Aristotle University of Thessaloniki.
}}

\maketitle

\begin{abstract}
\blfootnote{Copyright 2020 CERN for the benefit of the ATLAS Collaboration. \mbox{CC-BY-4.0} license.}
The Fast Tracker (FTK) is an ATLAS trigger upgrade built for full event, low-latency, high-rate tracking. 
The FTK core, made of 9U VME boards, performs the most demanding computational task. The Associative Memory Board 
Serial Link Processor (AMB) and the Auxiliary card (AUX), plugged on the front and back sides of the same VME slot, 
constitute the Processing Unit (PU), which finds tracks using hits from 8 layers of the inner detector. The PU works 
in pipeline with the Second Stage Board (SSB), which finds 12-layer tracks by adding extra hits to the identified tracks. 
In the designed configuration, 16 PUs and 4 SSBs are installed in a VME crate. The high power-consumption of the AMB, 
AUX and SSB (respectively of about 250 W, 70 W and 160 W per board) required the development of a custom cooling system. Even 
though the expected power consumption for each VME crate of the FTK system is high compared to a common VME setup, the 8 FTK 
core crates will use $\approx$ 60 kW, which is just a fraction of the power and the space needed for a CPU farm performing the 
same task. We report on the integration of 32 PUs and 8 SSBs inside the FTK system, on the infrastructures needed to 
run and cool them, and on the tests performed to verify the system processing rate and the temperature stability at a 
safe value.
\end{abstract}


\section{Introduction}
\label{sec:introduction}
\IEEEPARstart{T}{he} VME boards developed for the Fast TracKer Processor (FTK) \cite{b01}, an ATLAS \cite{b02} trigger upgrade, 
perform the most power consuming task of the system. The FTK is able to reconstruct, with high resolution, all of the tracks with 
transverse momentum above 1 GeV within short latencies (few tens of microseconds); it is compact (4 VME racks) and power 
usage is low ($\approx$ 60 kW). 

The FTK is organized in a pipelined architecture. A key role is played by high performance FPGAs, while most of the computing 
power is provided by full-custom ASICs named Associative Memories (AM) \cite{b03}. 

\begin{figure}
\centering{
$\vcenter{\includegraphics[width=2.0in]{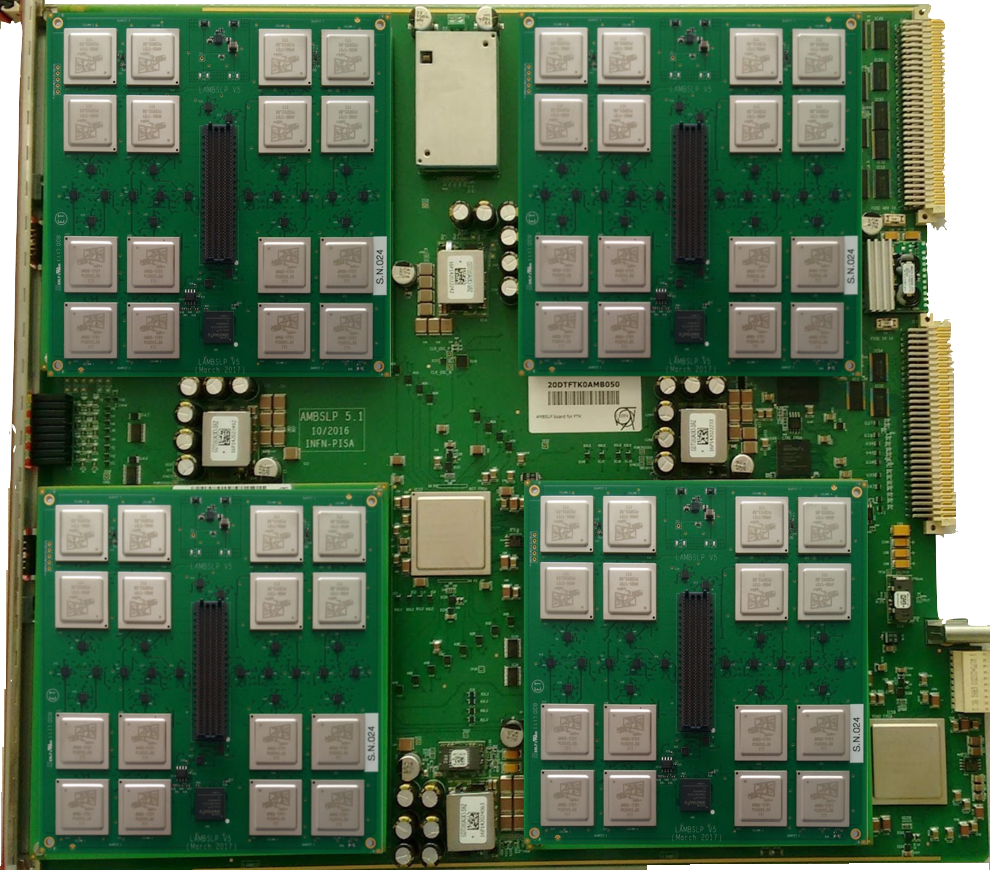}}$
$\vcenter{\includegraphics[width=2.0in]{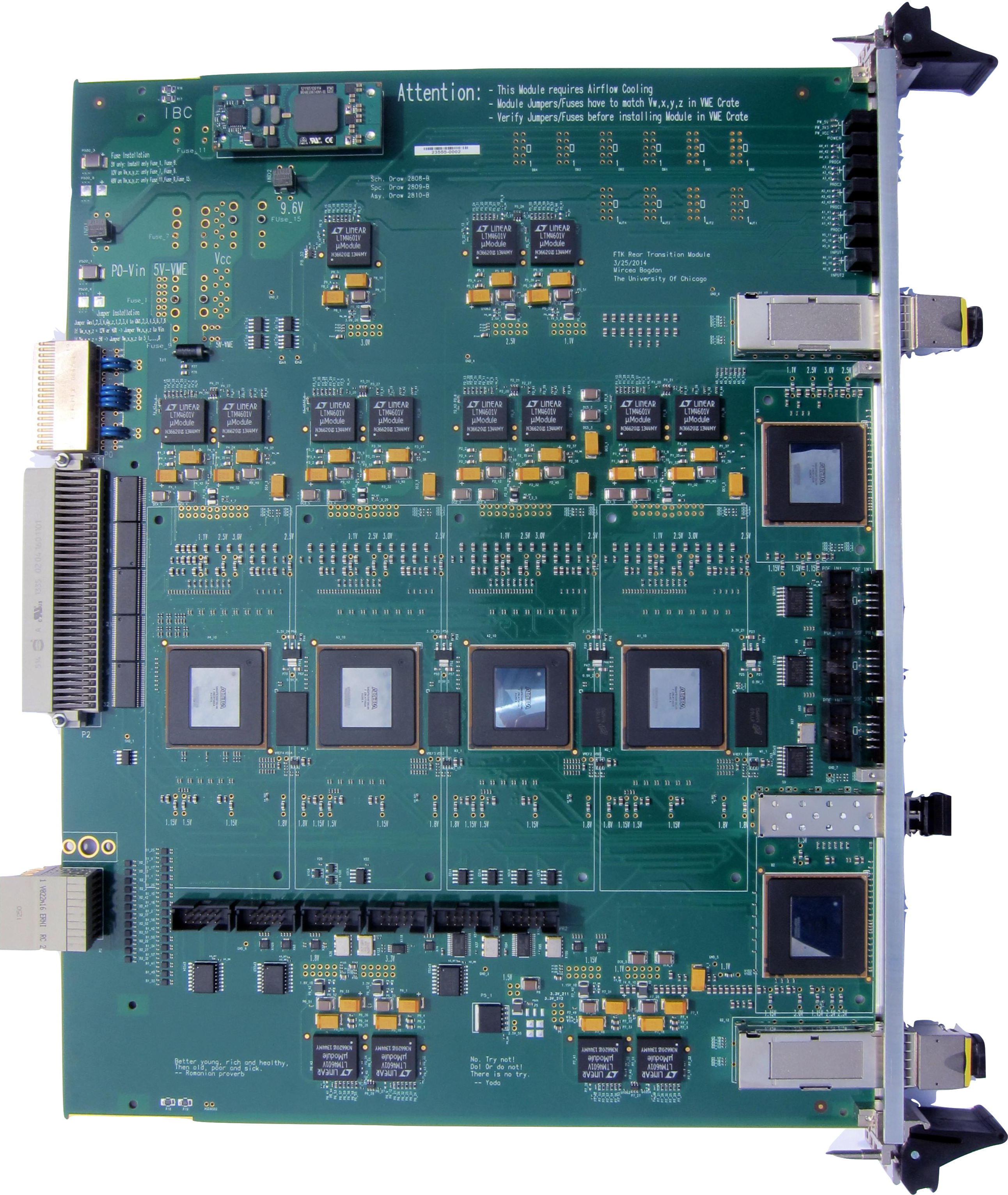}}$
}
\caption{ A PU is composed of one AMB (top) and one AUX card (bottom). In the AMB picture, it is possible to distinguish the 4 LAMBs 
and the 64 AM chips. The LAMBs in the picture show their internal face, in order to make it possible to see the AM chips.}
\label{fig1}
\end{figure}

The 9U VME cards inside FTK are the “Associative Memory Board Serial Link Processor” (AMB), the “Auxiliary card” (AUX) 
and the Second Stage Board (SSB). The AMB, full of AM chips, is managed by the large Rear Transition Module named the AUX 
card, full of powerful FPGAs. They sit in the same VME slot and constitute (see Figure \ref{fig1}) a Processing Unit (PU).  The PU 
is the FTK core that solves the huge combinatorial task with pattern recognition. It is also the most power and time 
consuming part of FTK. 

The AMB executes the low resolution pattern matching (PM) \cite{b04}, working on 8 detector layers. It is based on the use 
of a large bank of pre-computed patterns of trajectory points. The whole AM system stores 1 billion AM patterns. It 
recognizes track candidates at low resolution to perform the demanding task of tracking at the detector readout rate. 
The AUX validates or rejects the track candidates found by the AMB by performing high resolution track fitting (TF) \cite{b01}. 
The two functions PM and TF are executed in pipeline and interfaced by the Data Organizer, which provides reduced 
resolution hits to the PM and well-organized full-resolution hits to the TF. The PU output (8-layers tracks) is sent to the SSB, 
which adds 4 extra detector layers to produce 12-layer tracks.

The AMB is assembled with four mezzanines, the Little Associative Memory Boards (LAMBs), for a total of 64 AM chips 
per board. The infrastructure design for this compact and powerful system is described in Ref. \cite{b05}. 

Over 200 boards were produced, including 128 AUXs, 64 AMBs and 32 SSBs.
Two PUs and one SSB were included in 2018 ATLAS data taking. 
Two full VME crates (32 PUs and 16 SSBs) were used for cooling tests.

\section{ Timing tests of the Integrated VME boards – Monitoring Tools}
The monitoring of FTK-board dataflow is based on circular memories in the firmware called Spy Buffers\cite{b04}.  
Circular memories are continuously written with the data processed by the board unless a Freeze signal 
is asserted. When the memory is frozen, it can be readout through VME bus for error diagnosis or standard monitoring 
functions. The Freeze signal has been successfully used during the tests to detect rare errors. It is activated 
on error conditions and propagated to Freeze all of the involved spy buffers. 

Other quantities are monitored during data processing: (a) to understand pipeline bottlenecks the 
fraction of time each input FIFO is Empty or Half-full is measured; (b) the event processing rate; (c) the temperatures, 
I/O currents and power consumption for each power voltage; (d) specific tools built for the AMB. 
For each event, the time needed to load data into the AM chips and to send 
matched patterns to the output is measured for each I/O stream. These measurements provide data for the 
pipeline studies, which is crucial for ensuring the ability of the system to cope with the required latency.

In the AMB, clusters (Hits) are received in parallel on 8 S-LINKs \cite{slink} (HIT Time) and each Hit is distributed to all  
patterns with maximum fan-out: each group of 8 Hits reaches all patterns in a single clock cycle. The 
track candidates (Roads) found by the AM chips are readout in parallel on 16 S-LINKs (Road Time). While the AM chips 
receive an event, the previous event is readout from them in parallel. The End Event word 
(EE) declares the end of an event download or an event readout to/from the AM chips. 

The processing time is measured for all the 8 HIT links and the 16 ROAD links using the same firmware module, the Time 
Monitoring Module (TMM). The monitoring dedicated module measures the number of clock cycles between a “Start” and 
a “Stop” signal.

\begin{figure}
\centerline{\includegraphics[width=3.5in]{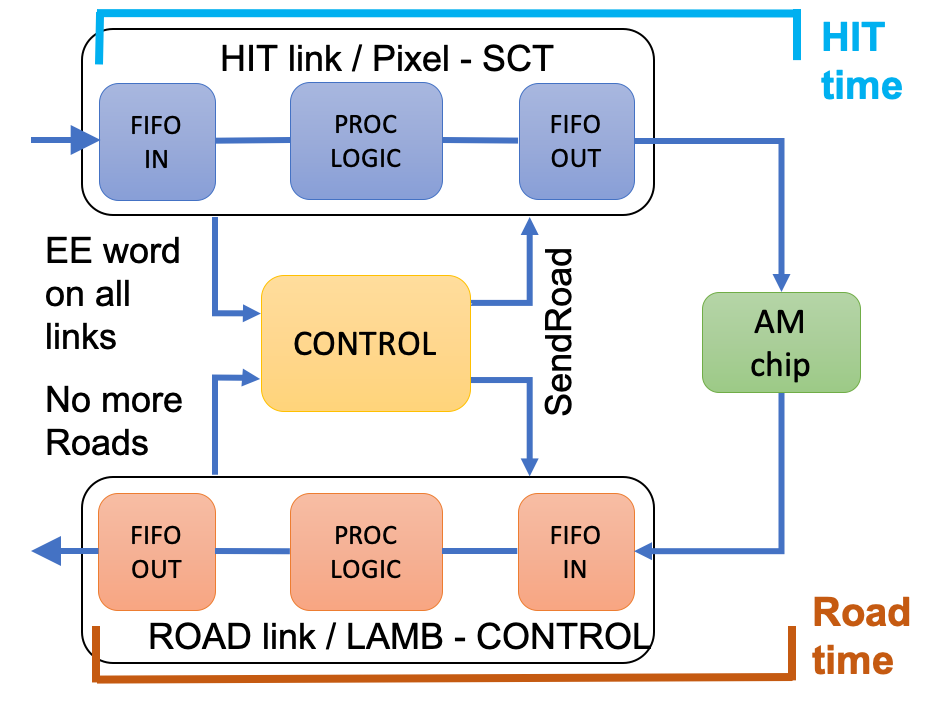}}
\caption{ Sketch showing the portion of the AMB HIT and ROAD link logic where the times are measured.}
\label{fig2}
\end{figure}

Figure \ref{fig2} shows the signals used to generate the “Start” and “Stop” signals of 
the counters for both the HIT and ROAD link types.
Both of the counters start at the activation of the Send\_Road signal sent by CONTROL to both the Hit and ROAD 
links, to start the downloading of a new event and the readout of the previous one. 
In the case of HIT links the time counter stops when the End Event word is detected in all the links. 
The ROAD counter is stopped when the End Event word has reached the OUT FIFO for each ROAD link. 

\begin{figure}
\centerline{\includegraphics[width=3in]{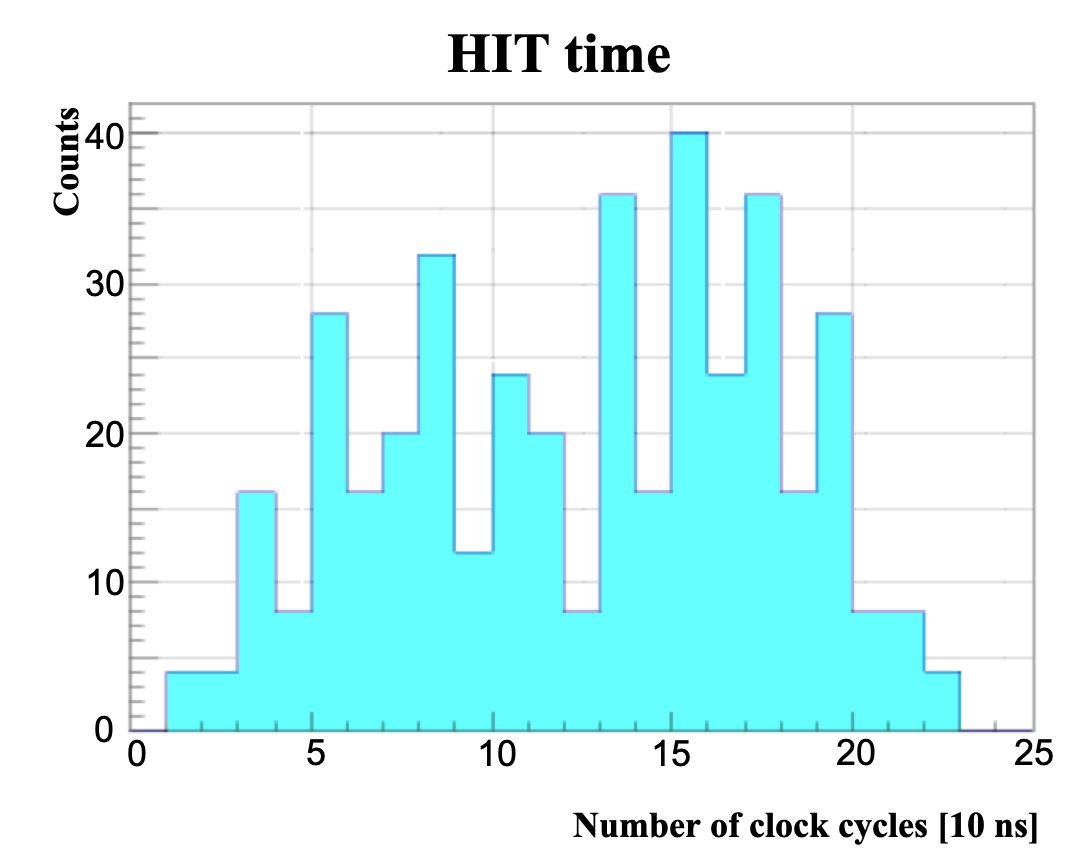}}
\caption{ Input link time distribution measured with simulated random events.}
\label{fig3}
\end{figure}

\begin{figure}
\centerline{\includegraphics[width=3.5in]{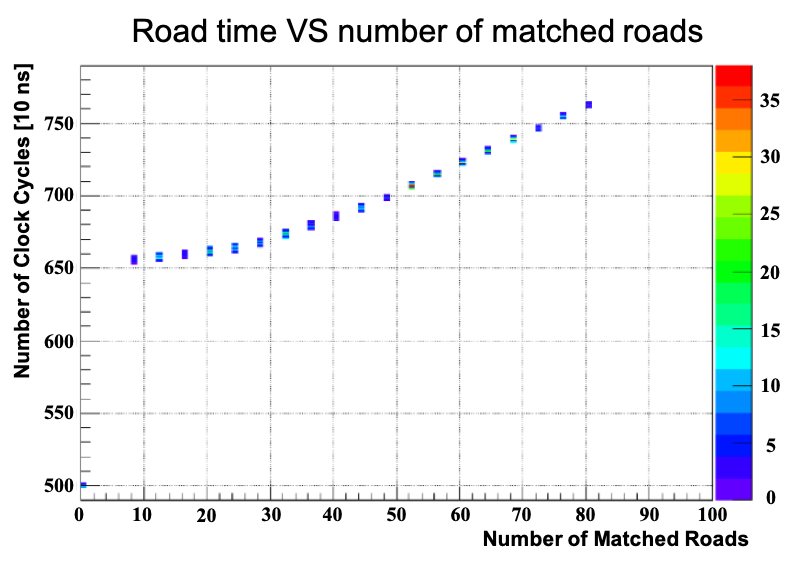}}
\caption{ROAD output link times as a function of number of matched roads.}
\label{fig4}
\end{figure}

The link timings have been measured with simulated random events. Figures \ref{fig3} and \ref{fig4} show the results of the 
time measurements for data transit in the AMB on each input (Figure \ref{fig3}) and output (Figure \ref{fig4}) event. The events 
 and the set of patterns stored in the AM chips were picked such that roads were constrained to fire in the position of the AM pipeline 
where the latency is maximum (worst case).

The input links have a time significantly smaller than the output links. The lowest measured time for the output links is around 
500 clock cycles, corresponding to events with no matched roads. In fact, the output chip was programmed 
to count 500 clock cycles at the AM chip output after the last received road before declaring ``Ended'' the event to avoid the loss of roads 
delayed inside the Associative Memory pipeline. The delay to get data through the daisy chain and ROAD output logic 
explains the extra clock cycles that separate empty and not empty events. 

These timing studies allowed us to significantly decrease the conservative 500 clock cycles count originally set for the End of Event 
identification, bringing the new value to 50 clock cycles. This change significantly reduces the output link times. 

In conclusion, these measurements show that the AMB latency meets the timing requirements for a 100 kHz event processing rate.

\section{VME boards power consumption}
Since the power consumption of the board components changes based on their configuration, a direct measurement is made 
rather than relying on predictions from the manufacturers. 
The resource utilization (in typical FPGA designs, some of the resources are not used after the configuration and thus they do not 
consume any power) and the switching activity (the number of signal transitions in a clock period) are the main 
contributors to the power variation. 

The power consumption of the AM chips also requires a careful measurement. In this element, the power consumption is 
not related to its configuration, but to the rate and composition of the data being processed.

Being FPGA-based, the power consumption of the AUX and SSB boards during operation can be considered quite stable. 
Therefore, the variations upon the different running conditions are small. 
The measurements have been performed by programming the two boards with the latest stable firmware (FW) versions (end of 2018), 
and running them in standalone (not connecting them to the up-stream/down-stream boards) while processing simulated data.
The AUX boards, running in loop mode on data directly loaded on the FPGAs FIFOs, have been measured to consume about 70 W.
The SSB boards, fed with data coming from an emulator connected to the board, have been measured to consume about 160 W. 
Since no relevant changes in the FPGA resources utilization or in the switching activities are expected, these values 
can be considered to be close to the operational ones.

The estimation of the power consumption of the AMB relies on its two different processing components: 
FPGAs and AM chips. While, as for the AUX and SSB case, the AMB FPGAs are characterized by an almost fixed power 
consumption (for a given FW version), the AM chips show drastic variations. As already mentioned, these variations are 
related to the characteristics of the data undergoing processing, and in particular to the input data rate and to the 
Bit Flip number (BF), i.e. the number of bits that flip in the AM memory lines during the pattern matching procedure. 
The total AMB power consumption can be, therefore, computed as:

\begin{equation}
\label{eq1}
AMB_{\text{power}}=AMB_{\text{core}}+4\cdot(LAMB_{\text{core}}) .
\end{equation}

where $AMB_{core}$ represents the power consumption of the AMB mother board alone (FPGAs component), while $LAMB_{core}$ 
represents the power consumption of the LAMB boards, which is data dependent (AM chip component). The $AMB_{core}$ value has been 
measured to be about 100 W. The $LAMB_{core}$ is in turn the sum of two different contributions.

The first contribution comes from the dummy hit value, a 32-bit word that is sent to the AM chip, alternated to the IDLE 
word, during the IDLE state to avoid a strong current reduction when no or too few real data are available. Because of 
the high data fan-out in the memory bank, the current on the 
AM chip net is not constant, but varies from a quiescent current of 20 A to a peak current of 220 A in a few hundreds of 
nanoseconds. A very powerful DC-DC converter is used to power each LAMB, but even with this device it is not possible 
to avoid the effect of the fast large peak of current, resulting in a drop of the voltage level. In order to mitigate this 
problem, a dummy word is sent to the AM chips while in IDLE, i.e. while it is not processing data, helping to increase the 
quiescent current and, thus, reducing the voltage drop. 
The dummy word doesn't fire any pattern on the chip, but can be chosen so that the number of bits that flip in the AM 
lines at each clock cycle can be tuned as desired, managing the IDLE power consumption of the chip. 
The choice of the value of the dummy hit is finally a compromise between the increase of the quiescent current of the 
chip net, requiring a big bit flip value, and the necessity of reducing the total power consumption of the board.

\begin{figure}
\centerline{\includegraphics[width=3.5in]{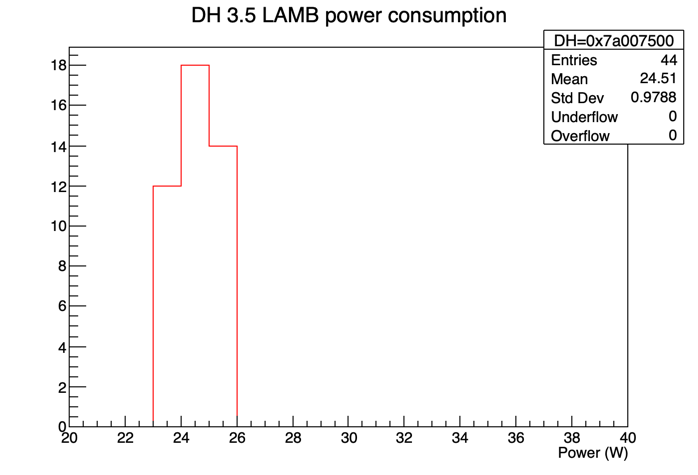}}
\caption{ LAMB power consumption for a value of dummy hit of 0x7a007500 (corresponding to a bit flip of 3.5). During this 
measurements no data were sent to the AM chips, the resulting power consumption is only due to the dummy hit.}
\label{fig6}
\end{figure}

The dummy hit value selected as optimal for operations corresponds to a BF value of $3.5$. In Figure \ref{fig6} the 
result of the measurements performed to characterize the IDLE power consumption using this dummy hit is presented, 
showing a value of about $25$ W.

The second component contributing to the LAMB power consumption is the mean number of bit flips inside the data and the rate 
at which these data are sent to the AM chips for comparison. From the expected FTK running conditions, 
the maximum value for the data word undergoing matching in the chips is represented by a 13-bits word. Considering continuous
data with random data words distributed flat between 0--8191 (using 13 bits), the power consumption of the LAMB has been 
measured to be about 48 W. This value has been measured feeding the AMB with continuous simulated data, ensuring no dummy hit 
contribution to the power consumption value.

In order to have a conservative estimation of the rate, the worst case scenario for the FTK running condition has been considered. 
From simulations, considering a pile-up value of $\mu = 57$, the maximum number of clusters among the different FTK regions and 
detector layers is expected to be 340 clusters/layer. This value was used for the computation of the power consumption, 
giving a $22\%$ margin over the mean value on the different detector layers. It is important to notice that the cluster number does not 
depend on the set of patterns stored in the AM chips, but only on the detector geometry and pile-up. Using a linear approximation for 
the scaling of the number of clusters with the pile-up, a number of 360 clusters/layer is considered for the \mbox{$\mu$ = 60} operation case.

Assuming a mean number of 360 clusters to be processed every 10 $\mu$s at $\mu$ = 60, and given a 100 MHz clock cycle for the AM chips, 
36\% of the cycles are spent processing data with a power consumption of 48 W. The remaining cycles are spent in IDLE 
mode with the board consuming the \mbox{25 W} given by the dummy hit. The resulting total AMB power consumption at a pile-up 
\mbox{$\mu$ = 60} is
\begin{equation}
\label{eq1}
AMB_{\text{power}}=100\:W + 4 \cdot (48\:W \cdot 0.36 + 25\:W \cdot 0.64) \approx 233\:W
\end{equation}
with a mean LAMB power consumption of about 33 W. 
The total VME rack power can be computed as the sum of the different board's powers, plus the custom fan tray system measured 
to consume 3.4 kW per VME rack.
In Table \ref{power}, a summary of the various VME components power consumption is shown. Considering a 90 \% efficiency for the 
power supply ($PS_{eff}$), the total VME rack power is:
\begin{equation}
\label{eq1}
\begin{split}
 &Rack_{\text{power}}= (((AUX_{\text{power}} + AMB_{\text{power}})\cdot16 +  \\ 
         & + SSB_{\text{power}}\cdot4) \cdot 2 + Fan_{\text{power}}\cdot2) / PS_{\text{eff}} \approx 16\:kW
\end{split}
\end{equation}
\begin{table} 
  \caption{Table summarizing the power consumption of the different FTK VME components of each FTK rack.} 
  \label{power}
	\setlength{\tabcolsep}{3pt}
  \centering
  \begin{tabular}{|l|c|c|}
      \hline
          FTK VME component  & Power (W)              & Component number \tabularnewline
      \hline
          AUX                & 70              & 32        \tabularnewline
      \hline
          SSB                & 160             & 8         \tabularnewline 
      \hline
          AMB                & 212-244  & 32        \tabularnewline
      \hline
          Fan tray 			     & 850            & 4         \tabularnewline
      \hline
          Full rack          & $\approx16000$           &                  \tabularnewline
      \hline
  \end{tabular}
\end{table}

The standard VME cooling, coupled with the standard rack water cooling system, is not powerful enough to dissipate 
a total power of $\approx$16 kW. A custom cooling system based on powerful fan trays has therefore been adopted.

\section{Temperature limits and cooling infrastructure}
Each board component has a given temperature range in which its safe operation is guaranteed. The maximum temperature 
threshold is generally provided by the manufacturer of the given component. Keeping the temperatures low during operation 
also increases the lifetime of the boards, reducing the risk of failures. Given the high 
number of FPGAs (about 2000 elements) and of ASICs (about 8000 AM chips) that compose the FTK system, reducing 
the component failure risk is of primary importance for the project itself.

For AUX and SSB the temperature limit is driven by the processing FPGAs. The AUX board is equipped with 6 commercial 
Altera Arria-V FPGAs \cite{AlteraV}. This FPGA is rated to an operational temperature limit of 85 $^{\circ}$C. No damages to the component are 
expected for temperatures up to \mbox{125 $^{\circ}$C}. The SSB board power consumption is driven by 4 Xilinx Kintex-7 FPGAs \cite{Xilinx:Series7}.  
Version 4 (V4) boards use commercial FPGAs rated to 85 $^{\circ}$C, and a second production (V5) use industrial FPGAs rated to 100 $^{\circ}$C.
The AMB is equipped with 4 FPGAs. The HIT and ROAD chips are commercial Xilinx Artix-7 FPGAs \cite{Xilinx:Series7}, rated to an operational 
temperature threshold of 85 $^{\circ}$C. The AMB data control logic is instead based on two Xilinx Spartan-6 FPGAs \cite{Xilinx:Spartan6}, 
that present 
the same temperature thresholds. For both FPGA versions, no damages are expected for temperatures up to 125 $^{\circ}$C.
Beside the FPGA limits, the most temperature sensitive elements are the AM chips. Being a custom element, an 
exhaustive data sheet is not available. Estimations have rated the AM chips to be damage-safe up to 100 $^{\circ}$C, but a 
conservative operational temperature threshold has been set to 80 $^{\circ}$C.

A protection system has been put in place to prevent possible damage from overheating. This system relies on two different 
components. 

The first component is a FW protection, able to immediately stop the board dataflow in case the measured 
temperature is found to surpass a given threshold. While the FPGA temperatures are directly measured by the 
components themselves, the AM chips measurement relies on temperature sensors placed around the AM chips, on 
the LAMBs. This protection stops the data sent to the AM chips in a whole LAMB, letting the temperature 
decrease. 

In addition to the FW component, the second one is a software (SW) protection, managed by the ATLAS DCS (Detector Control 
System \cite{{DCS}}) and able to shutdown a VME crate in case of problems. In general, the DCS is a highly distributed system that is 
used to supervise and monitor the ATLAS experiment infrastructure. The FTK DCS is used to control and monitor the FTK 
infrastructure, retrieving data such as the board temperature, the air temperature in the crate and in the fan trays, 
the fan speeds and the power supply parameters. In addition, the FTK DCS manages the cooling system 
in response to the retrieved data, as well as managing the PS emergency interlocks. Exploiting this, the DCS has 
been programmed to read the rack air temperature, and to trigger the PS interlock in case the read value surpasses 
a given threshold. The interlock shuts down the power to the crate, keeping only the power to the cooling system 
(the fan trays). The rack air temperature threshold at which the DCS triggers the shutdown was preliminarily set 
to 32 $^{\circ}$C but, as a result of the cooling tests, described later, was finally increased to 38 $^{\circ}$C. 

In order to dissipate a power of more than \mbox{$\approx 16$ kW}, six 1U heat exchangers are mounted 
in each VME rack. The heat exchangers are cooled by means of a cold water flow injected at 14 $^{\circ}$C, with a 
pressure $\Delta P=4.5$ bar.
These heat exchangers, in the given configuration, are proven to be able to dissipate 3.8 kW of power each \cite{rack_cooling}. 
The high number of heat exchangers needed increases the air resistance in the rack.
Together with the high board density (increasing the air flow resistance as well), this poses the requirement of powerful fans, able to help 
the cooling system by increasing the air flow and its uniformity between the different crate slots. 

Four custom-made fan trays, composed of nine powerful fans (able to run with a speed above 10.5 krpm) have been developed and 
produced. The choice of custom fan trays is driven by the fact that the commercially available ones do not  
provide enough air flow through the rack components, as demonstrated in studies performed during the development phase of 
the FTK VME cooling infrastructure \cite{b05}.

\begin{figure}
\centerline{\includegraphics[width=3in]{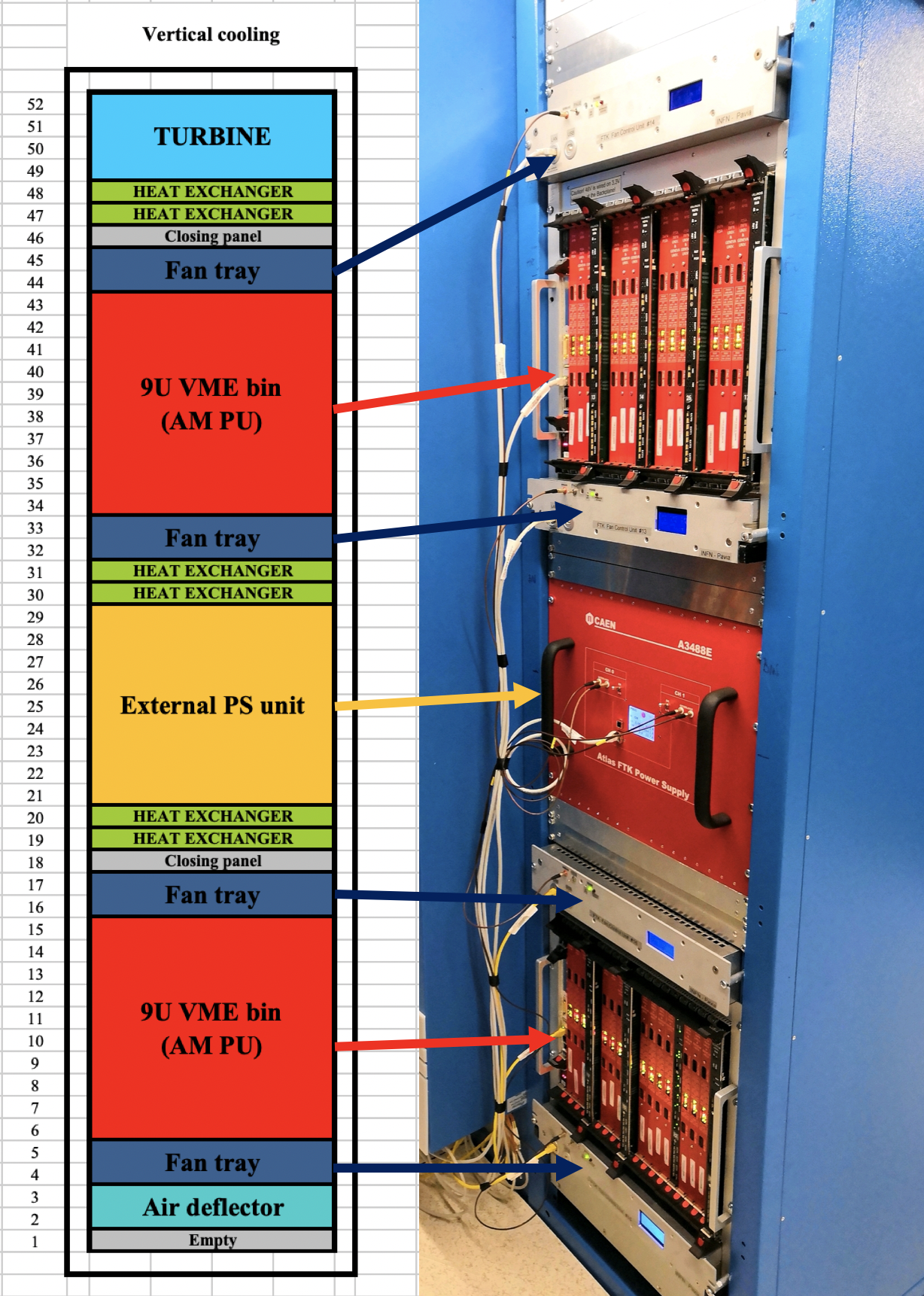}}
\caption{Sketch representing the VME rack configuration, together with a picture of a fully populated rack.} 
\label{VME_rack}
\end{figure}

In the final FTK VME cooling system composition, each VME rack is equipped with the six heat exchangers, the standard rack turbine, 
and with four of the custom fan trays, placed just under/above the crates. Figure \ref{VME_rack} shows a sketch of the final VME rack 
configuration, together with a real picture of a fully populated rack. 

The nine fans of each fan tray are organized into three rows of three fans each, two rows in the front of the rack, where the 
AMB and SSB boards are located, and one row on the back, where the AUX cards are placed. 
Electronic control and monitoring are implemented with an Arduino microcontroller, Mega 2560 type, equipped with a local 
display for every fan tray. Local/remote ON/OFF is available, and four speeds (40\%, 60\%, 80\%, 100\%) are settable 
independently within the three rows of fans. Six temperature sensors are positioned inside the fan tray to monitor the 
environmental temperature. The fan speed selection is controlled, through the Arduino, by the DCS system. The speed of the 
fans is automatically managed and increased/decreased by the DCS with the increase/decrease of the temperatures measured 
inside the fan trays.

For testing purpose only, a chiller, able to further cool down the water entering the rack existing cooling system (i.e. to the heat 
exchangers), was also provided for the FTK racks. 
A set of tests with this cooler water were performed, but the tests shown later demonstrate its use is not necessary.

\section{FTK VME Cooling tests}
In order to prove the performance of the cooling infrastructure, a series of cooling tests were performed, reproducing the 
board working conditions expected during the FTK operation.

\begin{figure}
\centerline{\includegraphics[width=3in]{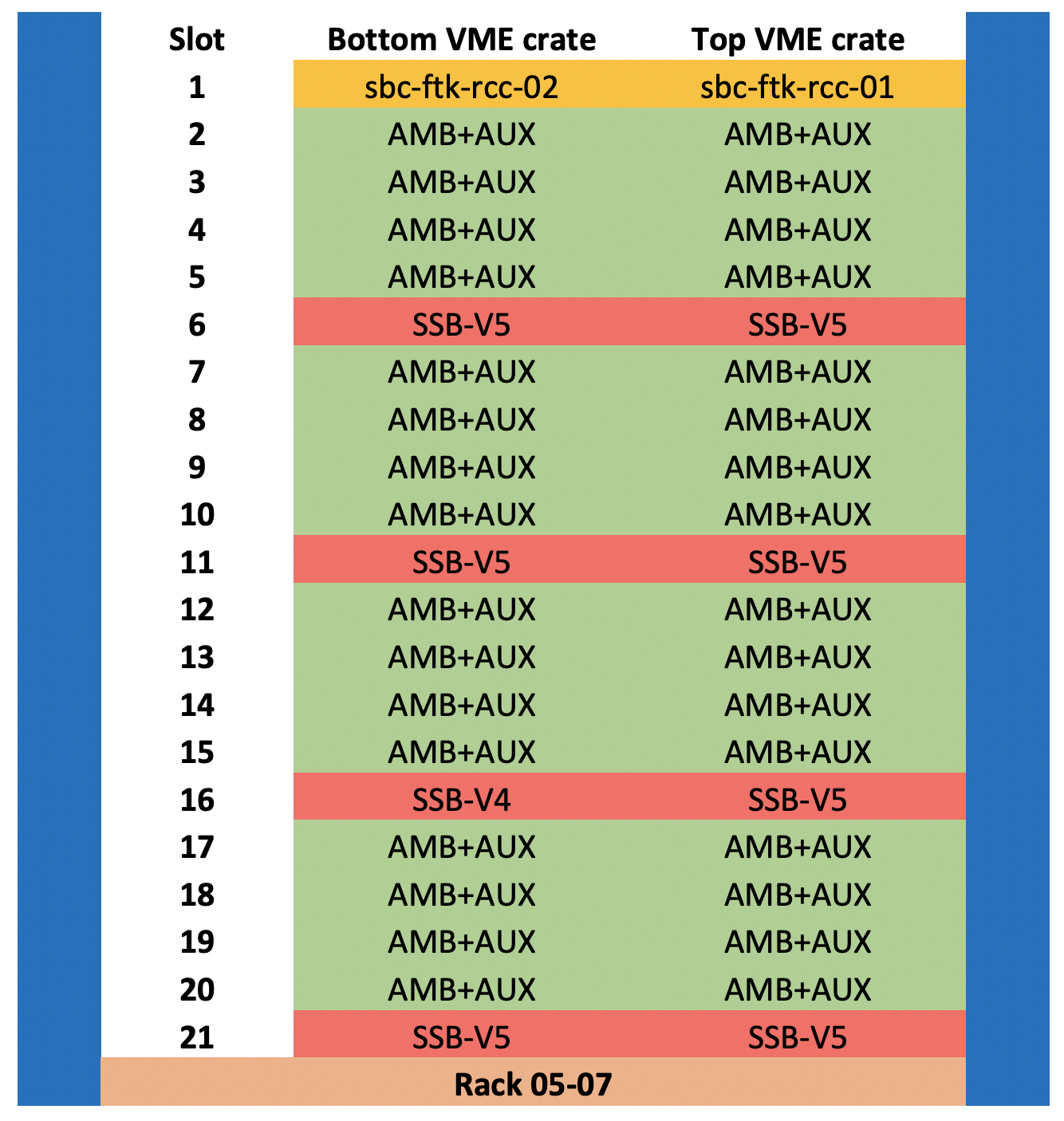}}
\caption{Sketch of the VME crates used during the cooling tests. The slots of each crate in which the 16 PUs and the 
4 SSBs are placed are shown. Only one V4 SSB was used during the tests, placed in slot 16 of the bottom crate.}
\label{VME_crate}
\end{figure}

The cooling tests have been performed using one of the VME racks fully equipped, with an FTK power supply, two VME crates fully 
loaded with boards, and the final version of the cooling system. Figure \ref{VME_crate} shows a sketch of the boards positions inside the 
testing crates.

\begin{figure}[t]
\centerline{\includegraphics[width=3.5in]{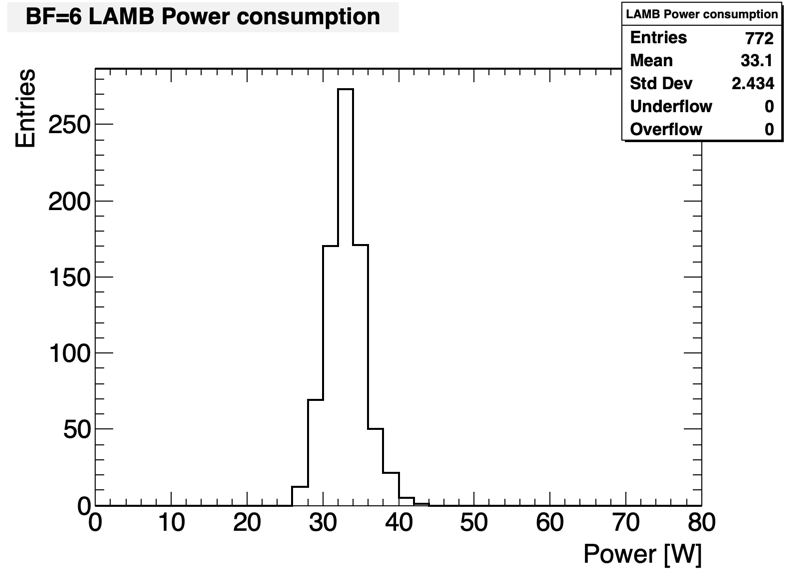}}
\caption{Distribution of the LAMB power consumption for the BF = 6 simulated data.}
\label{bf6_histo}
\end{figure}

For the tests, the AUX cards ran a loop over simulated data directly loaded in the input FIFOs. 
The power consumption of this board was about 70 W.
The two versions of SSB boards were used. However, only one V4 SSB was available for the tests. All the SSBs were running 
on data provided by an emulator, directly connected to the boards by the respective RTM cards. The power consumption of these 
boards was about 160 W. The AMB boards were run in loop mode on different sets of simulated data. The different data sets were 
produced with different predefined bit flip values (6, 6.5 and 7), in order to select a different power consumption of the AM 
chips for each of the measurements. 
For sake of brevity, only the tests performed with simulated data representing a BF value of 6, corresponding to the expected FTK
operation conditions, will be presented. In Figure \ref{bf6_histo}, the distribution of the LAMB power consumption for the testing 
data is presented. The measured 33 W per LAMB corresponds to the expectations presented in the previous section.

At the time of the tests, the board's temperature information was only accessible through custom command line tools, not 
integrated in the FTK online software. A python wrapper of these tools has been developed and used for the measurements.
This wrapper was able to read directly from the boards' FPGAs the temperature registers and to perform 
the conversions required. Each measurement was performed in intervals of about 300 s. In order to have a direct measurement of 
the AMB power during the tests, the AMB DC/DC converters values of current and voltage were also measured. This was not required 
for AUX and SSB, which show stable power consumption.

The temperature of the fan trays and of the rack air was also measured. The data in this case were directly retrieved from the 
DCS. The DCS was also used to provide the values of the power provided during the tests by the rack power supply. This 
information was useful in order to perform a cross check between the power expected from the AUX and SSB boards and the 
measured power consumed by the AMBs.

Moreover, during the tests both the DCS safety system and the automatic fan speed control were disabled. The former was due to a
temperature threshold initially set too low for the most demanding 
tests. The latter was triggered to study the temperature profile of
the boards on different values of the fan speeds, that were set manually to a different value on each test.

\begin{figure}
\includegraphics[width=3.5in]{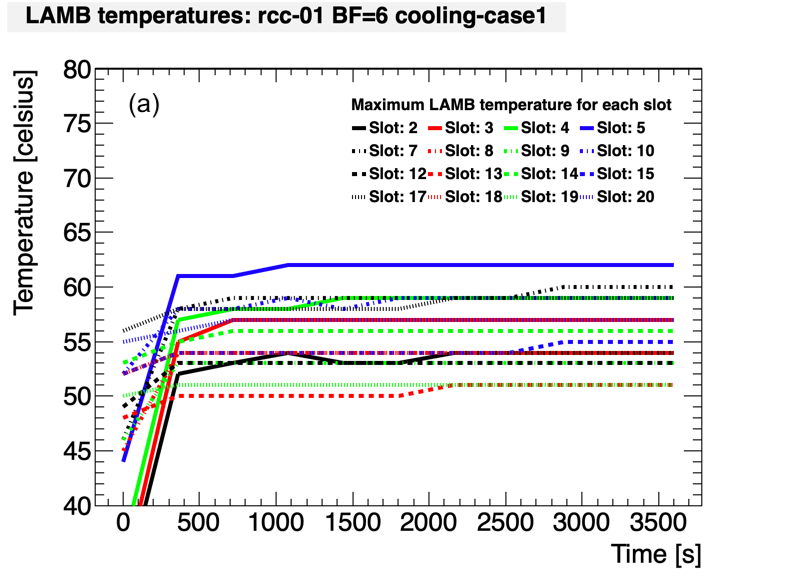}
\includegraphics[width=3.5in]{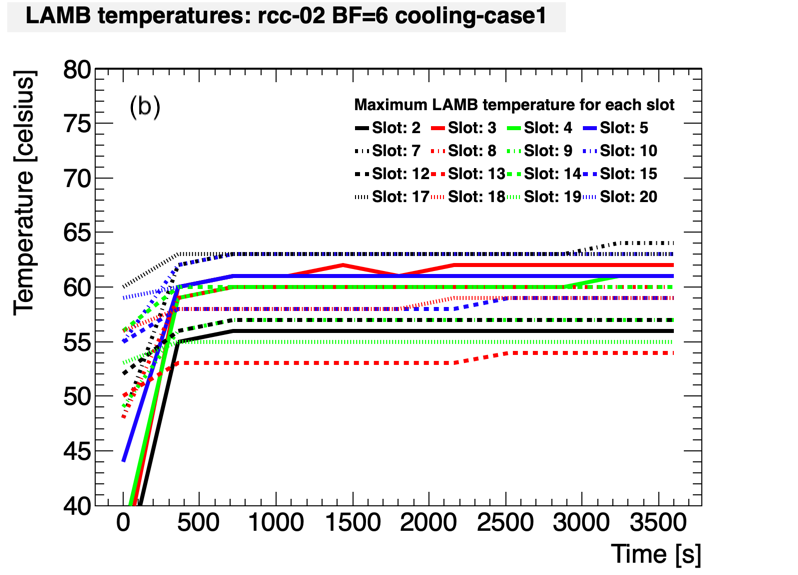}
\caption{Temperature profile of the hottest LAMB of each AMB for the top VME crate (a) and bottom VME crate (b). 
The AMB operational temperature limit is 80 $^{\circ}$C. In this configuration all the temperatures are well under control.}
\label{AMB_temp}
\end{figure}

Figure \ref{AMB_temp} shows the temperature profile of the hottest LAMB of each crate slot, for the upper (a) and lower (b) crates.
From both plots it is possible to note an initial rapid increase of the maximum temperatures, due to the heating up of the 
boards. This increase stops just after the second measurement, where the plateau is reached. In this configuration, the temperatures 
are well under the operational temperature limit.
From the comparison of the two crate results, the upper crate temperatures are slightly lower than the lower crate ones. This is 
expected due to the position of the heat exchangers inside the rack. As can be seen in Figure \ref{VME_rack}, only two heat 
exchangers will cool down the air coming from the upper VME crate, while four of them are placed between the lower and the upper crates. 
A solution to this problem would be to move one of the heat exchangers that are surrounding the PS to the very bottom of the rack, 
just after the air deflector. Since the difference in the temperatures between the two crates is small, this is not 
considered an issue.

\begin{figure}
\includegraphics[width=3.5in]{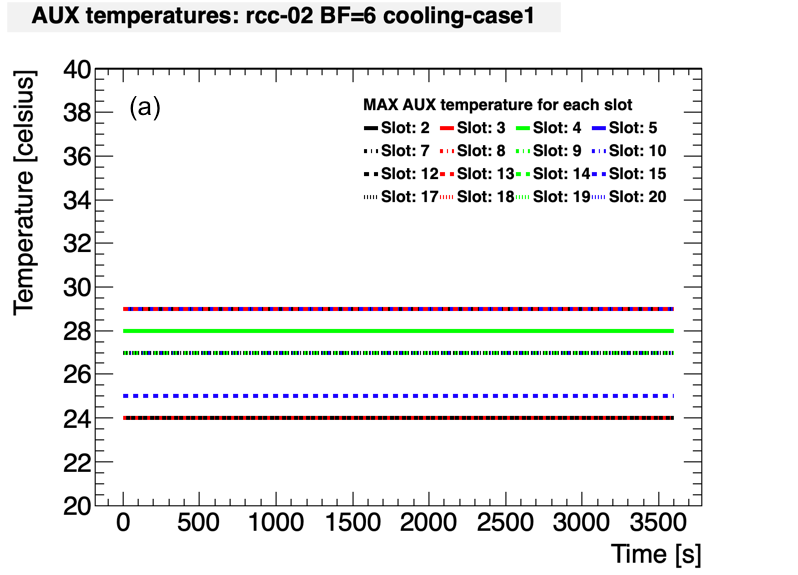}
\includegraphics[width=3.5in]{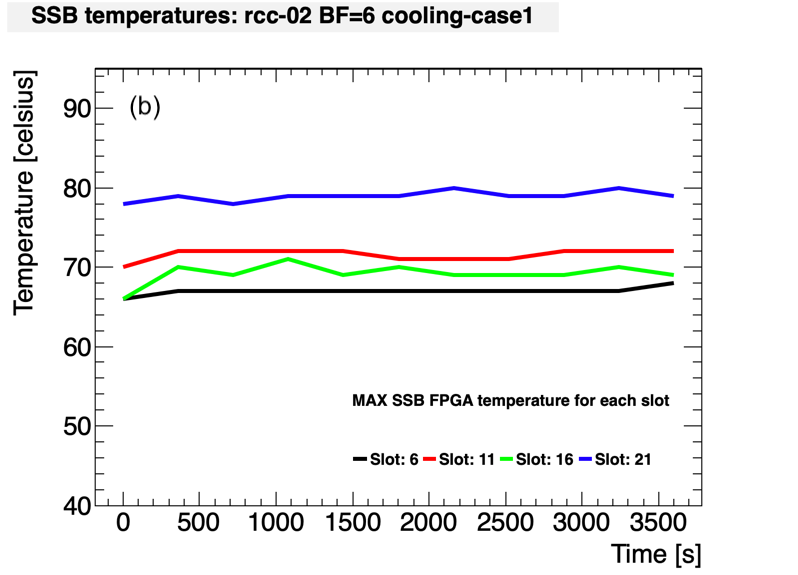}
\caption{ Temperature profile of the hottest FPGA of each AUX (a) and SSB (b) for the lower VME crate.
The slot 16 of the crate is equipped with the only V4 SSB used during the tests. All the temperatures are well under control.} 
\label{AUX_temp}
\end{figure}

Figure \ref{AUX_temp} shows the same measurements performed on the AUX cards (a) and on the SSB boards (b). 
For the sake of brevity, only the lower crate measurements (the hottest crate) are presented. 
Due to the low power consumption of this kind of boards, the temperatures are very stable. No increase is seen in 
the AUX temperatures because the first measurement is only made after the AMBs, so only the plateau is seen. 
The AUX cards are in the back of the crates, and therefore do not directly see the heat from the AMB and SSB.

The SSB the measurements were performed after the AMB ones, thus, only the plateau can be observed. The single 
V4 SSB used in the test is placed in slot 16 of the lower crate. The SSB placed in slot 21 shows temperatures of about 
\mbox{6 $^{\circ}$C} higher than the other boards. This is due to a non perfect distribution of the air flow inside the crate.

\begin{figure}[t]
\centerline{\includegraphics[width=3.5in]{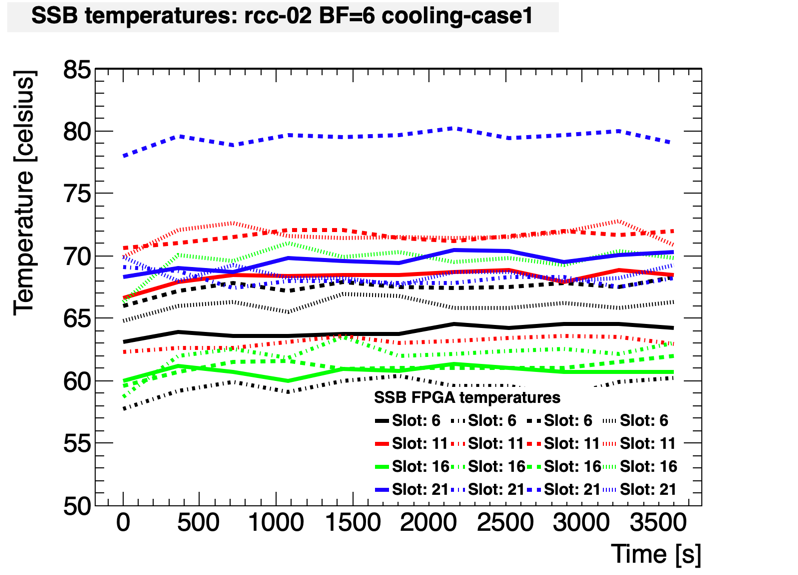}}
\caption{ Temperature profile of the four FPGAs of each SSB for the bottom VME crate. In the plot, each dashed line correspond 
to a different FPGA. The FPGA number of which each line correspond is the same as its order of appearance in the legend 
(e.g. continuous line: FPGA 0).} 
\label{SSB_temp}
\end{figure}

To better understand this problem, Figure \ref{SSB_temp} shows the same test measurements, but including the temperatures of all  
four SSB FPGAs rather than only the hottest one. In all the tests performed, the hottest FPGA of each crate has always been
the number 2 of slot 21. This FPGA is placed on the top back corner of the SSB, the one most difficult to be reached by the 
air flow produced by the fans. The FPGAs numbered 2 and 3 are usually hotter because they are directly above the two lower 
FPGAs, numbered 0 and 1. Between the lower FPGAs, the one placed in the back 
of the board is usually hotter than the front one. These results inform the final position of the boards in the crate 
as not much can be done to improve the airflow. A V4 SSB, for example, will not be
placed in slot 21, where the 80 $^{\circ}$C measured are very close to the operational temperature limit for that board.
In summary, all the measured temperatures are well under the operational temperature limit, for all the board types.

\begin{figure}
\centerline{\includegraphics[width=3in]{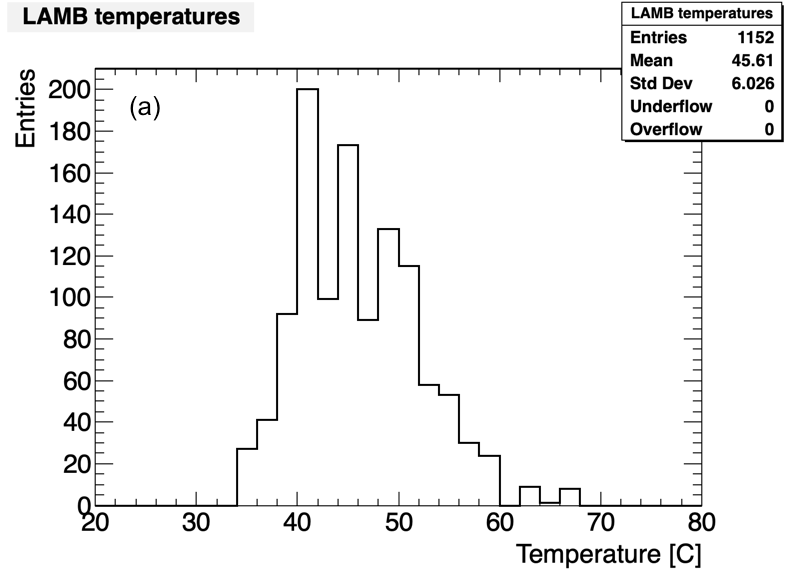}} 
\centerline{\includegraphics[width=3in]{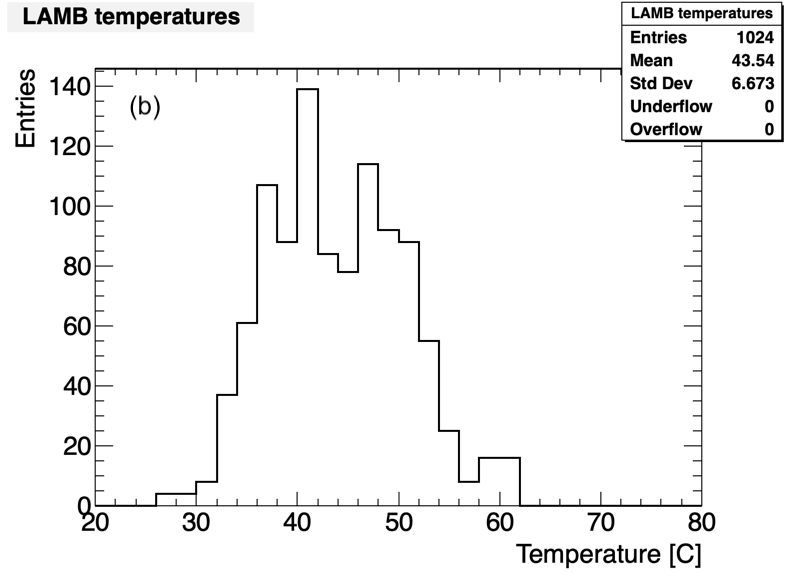}}
\caption{Distribution of the LAMB temperatures running the AMB with BF = 6 simulated data for two different racks (a) and (b). 
The VME crates on each rack were equipped with 8PUs and 1 SSB. A variation of 2 $^{\circ}$C has been observed between the two racks.} 
\label{rack_temp}
\end{figure}

Since all the tests were performed in a single rack, a second set of tests have been performed to study for possible rack 
to rack variations. Due to the lack of available boards, the test was performed with only eight PUs and one SSB per crate. 
In order to have coherent results, the same rack configuration was used performing the measurements on two different racks.
Figure \ref{rack_temp} shows the distribution of the LAMB temperatures for the two different racks tested. The measured mean 
LAMB temperature of the new rack is 2 $^{\circ}$C higher than the one of the 
standard rack. In order to be conservative, a variation of \mbox{5 $^{\circ}$C} has been considered as rack to rack variation for 
the final considerations on the cooling test results.

As a concurrent step to the cooling tests, optimization studies for both the best fan speeds configuration and the rack air temperature 
tresholds have been performed.

Varying the fan speed can cause differences in the air flux inside the racks, and it can increase/decrease turbulence. 
Moreover, reducing the speed of the fans can increase their lifetime. Tests, changing each time the settings of the fan 
speeds, have been performed in order to study the best configuration.
The tests were performed in the rack used for the cooling tests, using 16 PUs and 4 SSBs in each crate and using the speed 
configurations (fans at the top-bottom of the crate) \mbox{100\%--100\%}, \mbox{80\%--100\%}, \mbox{60\%--100\%}, 
100\%--80\%, and 100\%--60\%. The tests were also repeated for different AMB power consumption levels.

In all the tests performed, the lowest temperatures were measured in the fan speed configuration \mbox{100\%--60\%}. Running the bottom 
fan tray at a reduced speed with respect to the top one helps to avoid the creation of turbulences and to increase 
the uniformity of the air flux in the crate volume. Moreover, reducing the fan speed would significantly increase its lifetime. 
For this reason, the final fan speed of the FTK lower fans has been fixed to 60\%, leaving the speed of the top fans to be 
adjusted automatically by the DCS. In this case its speed will vary from the lower \mbox{60\%} level to the maximum \mbox{100\%} level as 
function of the temperatures read.

Finally, for the rack air temperature threshold optimization, tests running the boards above the expected power consumption levels
(in order to get closer to the board temperature thresholds)  
have been performed. For these tests, the fully populated rack used in the cooling test has been used, running the AMB with
BF = 7 simulated data, corresponding to a power consumption of 38 W per LAMB.
The maximum value of the air temperature obtained in this tests was 36 $^{\circ}$C. The maximum corresponding temperature 
for the LAMBs was 70 $^{\circ}$C. Since the test running condition is expected to be higher than the maximum condition 
expected for the FTK operation, the selected threshold for the SW rack safety system is 38 $^{\circ}$C. This would leave 
some margin over the maximum air temperature measured, still keeping the boards in a safe working condition.


\section{Conclusions}
Two FTK VME crates full of 16 PUs and 4 SSBs each have been successfully installed, and the system performance has been tested 
using simulated data. Even though the complete FTK processor’s power consumption ($\approx$ 60 kW) is a fraction of its CPU 
equivalent, its power density is much higher. This results in an unusually large power consumption per VME crate and
the need for a complex custom solution in order to ensure sufficient cooling. Extensive studies have been performed to investigate 
the system power consumption in real running conditions, and the ability of the developed cooling infrastructure to cope with the 
big power densities to be dissipated. Thanks to these studies and optimizations, the cooling infrastructure is proven to 
sufficiently cool the FTK processors under the worst case scenario. 
Moreover, the processing latencies were reduced based on the timing studies performed.  
The measurements performed demonstrate the timing requirements for a 100 kHz event processing rate are met. 

\section*{References}

\def\refname{\vadjust{\vspace*{-1em}}} 

\end{document}